\\

Title: Measurement of Krypton-85 in samples of atmospheric with Quantulus 1220 device without using a liquid scintillator

Authors: E.G. Tertyshnik, V.N. Ivanov, E.S. Abduragimov

The possibility of measuring the $^{85}$Kr activity with a liquid scintillation spectrometer (LSS) by registering the scintillations in krypton was investigated. It was found that at partial pressure of krypton less than atmospheric, the conversion efficiency of krypton scintillators amounts to 3 keV/photon and is 20 times lower than the conversion efficiency of liquid scintillators. The amplitude of the pulses due to scintillations in krypton is approximately equal to the amplitude of the pulses from Cherenkov radiation which occurs in the walls of the vial under the influence of $^{85}$Kr beta particles.

\\

# Measurement of Krypton-85 in samples of atmospheric with Quantulus 1220 device without using a liquid scintillator

E.G. Tertyshnik, V.N. Ivanov, E.S. Abduragimov ('SPA 'Typhoon', Obninsk, Russia)

The possibility of measuring the $^{85}$Kr activity with a liquid scintillation spectrometer (LSS) by registering the scintillations in krypton was investigated. It was found that at partial pressure of krypton less than atmospheric, the conversion efficiency of krypton scintillators amounts to 3 keV/ photons and is 20 times lower than the conversion efficiency of liquid scintillators. The amplitude of the pulses due to scintillations in krypton is approximately equal to the amplitude of the pulses from Cherenkov radiation which occurs in the walls of the vial under the influence of $^{85}$Kr beta particles.

**Key words:** atmosphere, Krypton-85, scintillation, Cherenkov effect, liquid scintillation spectrometer, conversion efficiency, beta particle.

INTRODUCTION

The long-lived radionuclide $^{85}$Kr (half-life of 10,75 years) is released into the atmosphere mainly as a result of emissions from spent nuclear fuel reprocessing plants. At present, the activity of $^{85}$Kr in the atmosphere is about $5\cdot10^{18}$ Bq or 135 mCi [1]. Increase in the $^{85}$Kr activity in the atmosphere can change electrical conductivity of the air through the ionization and cause geophysical effects which are difficult to predict. The content of this radionuclide in the atmosphere is regularly determined by a number of observatories located in Western Europe, Japan and Australia [2,3]. The applied laboratory installations separate krypton from the atmosphere using sorption of air samples with activated charcoal at a temperature of liquid nitrogen, following controlled desorption, during which the most part of sorbed nitrogen and oxygen is removed from the coal and finally the selection of krypton is made using a gas chromatograph [4,5]. Helium is used as a gas carrier. Since $^{85}$Kr nuclei decay by beta emission, the activity of $^{85}$Kr is usually determined by the introduction of the separated sample krypton into a gas-discharge counter. 1 m$^3$ of air contains only 1,14 ml krypton, so for the separation of a few ml of krypton the installation must rework 5 – 10 m$^3$ of air.

To decrease the necessary volume of the air sample from which krypton is extracted the authors of the paper [6] used a liquid scintillation counter (LSC) for the measurement of $^{85}$Kr activity in the krypton samples that were obtained at the outlet of the chromatographic column. The background of the used LSC (0,16 counts per second, the volume of the measured sample is 100 ml) is sufficiently lower than the background of gas-discharge counters and the detection efficiency of beta particles emitted by $^{85}$Kr nuclei is approaching 100%. However, the procedure of preparing the liquid scintillator and dissolution of krypton sample in it is very complicated [6].

This paper investigates the possibility of measuring the activity of krypton samples of a few ml volume with the help of LSC Quantulus 1220 without using a liquid scintillator by detecting the scintillations that arise in the gaseous krypton by $^{85}$Kr beta particles.

Previously it was shown [7,8] that at the pressure close to the atmospheric the exposure of ionizing radiation at the atoms of argon, krypton and xenon is accompanied by the appearance of photons, wavelength of which is corresponding to the vacuum ultraviolet region (for Kr - 150 nm). These quanta, in turn, can interact with the noble-gas atoms and cause the appearance of photons, the spectrum of which is in the sensitivity range of PMT radiometric installations.

EXPERIMENT

The krypton samples were placed into the 20 ml vials (flasks) made of low potassium glass (Perkin Elmer, USA) that are designed to measure tritium and carbon-14 using Quantulus 1220. Besides, the vials of the same volume made of organic glass (polymethylmethacrylate) in the laboratory were applied, Fig. 1. A standard glass vial (Fig. 1a) was closed with plug 1, at the center of which there was an opening 1,5 mm in diameter. That opening and gasket 2 from vacuum rubber 5 mm thick allowed to introduce the krypton sample into the vial through a syringe needle. Gasket 2 was glued to the inner side of the plug. A layer of epoxy provided a tight connection of the plug and gasket with the glass vial 4. To join the parts of the vial made of organic glass (Fig. 1b) dichloroethane-based glue was used. Air-tightness of the vials was controlled by the weighting on analytical scales. The mass of the vial with air and the mass of the vacuumized vial were compared and the weighting was repeated in several days. In case of leakage detection the vials were rejected.

Krypton and helium were introduced into the vials through the syringe needle in the following sequence: vacuumizing of the vial through the syringe needle up to pressure of 0.2-0.4 kPa was conducted. A piece of the pipe made of vacuum rubber 5x5 mm size 1 m length with the tap at the end was fixed to the output coupling of gas pressure regulator which was attached to the krypton ballon. When the tap was opened the pipe and the regulator were washed with krypton for air displacement; after that the tap was closed. The pipe side was pierced with the needle (the syringe plunger is at the zero position), under the overpressure krypton filled the syringe volume lifting the plunger. The needle was removed and the movement of the plunger fixed the certain volume of krypton in the syringe, for example, 5 ml. The syringe needle was introduced into the vial through the opening in the plug and the gasket, and due to the vacuity in the vial the syringe plunger moved to the zero position, and thereby 5 ml krypton completely moved into the vial. The needle was removed and then this krypton portion was diluted with 15 ml helium through another syringe needle. In this case, the total volume of mixture was equal to the volume of the vial – 20 ml. If 2 ml krypton was placed into the vial then the volume of helium attached to the vial was 18 ml. Therefore, the gas sample collected at the output of chromatography column was simulated (gas carrier – helium). After filling the vials, the opening in the plug was closed with vaseline.

Spectrometric measurements with LSC Quantulus 1220 of krypton samples prepared in such a way showed that the pulses of $^{85}$Kr beta particles are registered at the channels from 50 to 200, Figure 2. In these channel intervals the pulses from tritium beta particles are registered (average energy – 5,69 keV) in our laboratory during the routine measurements of hydrosphere samples prepared using the following cocktail: sample – liquid scintillator in a 8 : 12 ratio. A liquid scintillator OptiPhase HiSafe 3 of Perkin Elmer, USA is applied.

In assessing the quantum effectiveness of krypton-scintillator, the contribution of photons appeared in the interaction between $^{85}$Kr beta particles and walls of the vial should be taken into account. Figure 3 presented the dependence of electron energy threshold which is required for the generation of Cherenkov radiation in the medium on the light refraction index [9]. It is seen that Cherenkov radiation in the glass (index of refraction is 1,49) is occurred if the electron energy exceeds 180 keV. Since the maximum energy of $^{85}$Kr beta particles is equal to 687 keV, the significant part of the impulses registered with LSC may be the result of Cherenkov radiation. To determine this part the krypton samples were placed into the vial, the inner walls of which were covered with thin (1,6 mg/cm$^2$) aluminized film. The film didn't pass the photons induced by krypton scintillators and beta particles energy didn't decrease significantly during the film transiting. As a result, the separated effect was caused only by Cherenkov radiation.

From the table, which demonstrates the measurement results, it follows that the contribution of Cherenkov radiation to the total count rate is approximately 1/3 and 2/3 is made by pulses caused by krypton scintillations.

$^{85}$Kr activity given in the Table was determined considering the krypton volume in the vial and its specific activity. For the experiments the pure krypton (0,99999) with volume activity $^{85}$Kr 1,35 Bq/ml at normal temperature and pressure (0°C and 760 mm Hg) was used. During the introduction of krypton samples into the measuring vials (22 °C and 750 mm Hg) the $^{85}$Kr volume activity was calculated by Clapeyron-Mendeleev equation and amounted to 1,23 Bq/ml.

The obtained results suggest that in case of using for the measurements the vials made of organic glass the count rate for blank activity (the vial without krypton) will be 5 times lower than blank count rate in case of glass vial using and figure of merit, i.e. the ratio of efficiency square to the background turned out to be significantly higher for the plastic vial than for the glass vials. Lowering in the background by measuring the samples in plastic vials in comparison with glass vials is resulted from the reduction in the probability of background gamma-quantum interaction with the plastic due to its lower effective atomic number and density (approximately 2 times) compared to the glass vials.

In the last column of the table the values of efficiency of $^{85}$Kr beta particles registration are presented. These values were calculated by dividing the count rate of the pulses (except the blank count rate) by the number of beta particles emitted per second. For the krypton samples placed into the glass vials the average value of efficiency was 2,9 and for the samples placed into the vials made of organic glass – 2,7 pulses per beta-particle. Consequently, the detection efficiency of krypton-scintillator is 2 pulses per beta-particle and the expanded combined uncertainty is ± 1 pulse/beta for 95% probability. Calculation of the expanded combined uncertainty was executed taking into account the random and systematic component, and the latter was determined by the error of volume activity of used krypton ± 20%. Since the pulses from tritium beta particles, the average energy of which is 5,69 keV, are registered in the same interval of channels as the pulses caused by $^{85}$Kr and the average energy of $^{85}$Kr is 252 keV, so the conversion efficiency of the krypton-scintillator is approximately 20 times lower than the liquid scintillator efficiency. If the conversion efficiency of the liquid scintillator is 150 eV/ photon (if average energy of photon is 3 eV) then this value for krypton scintillator is approximately 3 keV/ photon.

The appearance of several pulses at the outlet of LSC in case of interaction with the gas scintillator of one beta-particle can be explained by excitation of metastable levels lifetime of which is longer than the resolving time of Quantulus 1220 installation.

CONCLUSION

The possibility of $^{85}$Kr activity measuring with Quantulus 1220 device in the krypton samples separated from the atmosphere without using a liquid scintillator has been demonstrated.

It has been established that at krypton partial pressure less than atmospheric, the conversion efficiency of krypton scintillators amounts to 3 keV/photon and is 20 times lower than the conversion efficiency of liquid scintillators and the amplitude of pulses due to the scintillations in krypton is approximately equal to the amplitude of pulses from Cherenkov radiation which occurs in the walls of the vial under the influence of beta particles.

It is experimentally shown that as a result of the impact of beta particles occurred by $^{85}$Kr decay on the gaseous krypton Quantulus 1220 device registers more than one pulse per beta-particle. It is possible if krypton atoms jump at the metastable levels lifetime of which is longer than the resolving time of used LSC.

The authors are grateful to Y.I. Baranov and A.A. Volokitin for the assistance in conducting the experiments.

**Table. Results from the analysis of krypton samples using LSC Quantulus 1220**

| Volume of krypton in the vial, ml; (material of the vial) | Count rate, puls./s | | Krypton activity in the vial, Bq | Efficiency, puls./beta-particle |
|---|---|---|---|---|
| | Observed | Except the blank count rate | | |
| 0; (glass) | 4.97 | 0 | 0 | - |
| 3; (glass) | 17.21 | 12.24 | 3.7 | 3.3 |
| 5; (glass) | 20.62 | 15.7 | 6.2 | 2.5 |
| 10; (glass) | 41.53 | 36.6 | 12.3 | 3.0 |
| 15; (glass) | 58.25 | 53.3 | 18.5 | 2.9 |
| 5, (glass + aluminized film) | 10.4 | 5.43 | 6.2 | 0.88 |
| 0; (organic glass) | 1.05 | 0 | 0 | - |
| 5; (organic glass) | 16.2 | 15.2 | 6.2 | 2.5 |
| 10; (organic glass) | 35.5 | 34.5 | 12.3 | 2.8 |

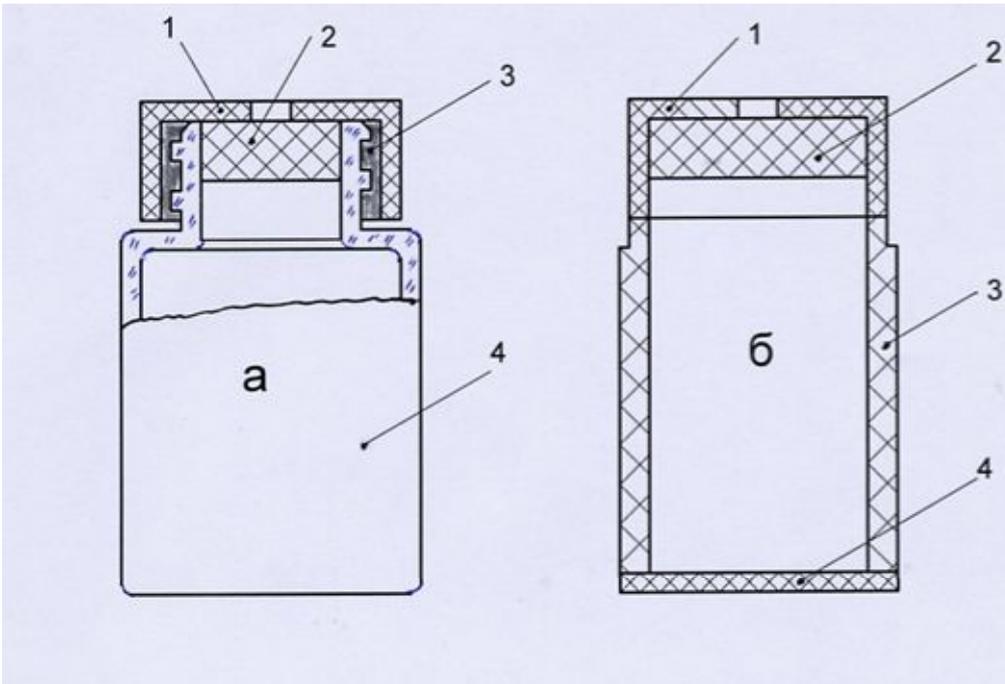

Fig. 1. Vials in which the samples of krypton were placed for the measurements using LSC
а − Vial made of glass, Perkin Elmer, USA:
1 − plunger made of organic glass; 2 − gasket from vacuum rubber; 3 − epoxy; 4 −vial made of low potassium glass.
б −Vial from organic glass made in the laboratory:
1 − plunger; 2 − gasket from vacuum rubber; 3 − feedwell; 4 − bottom.

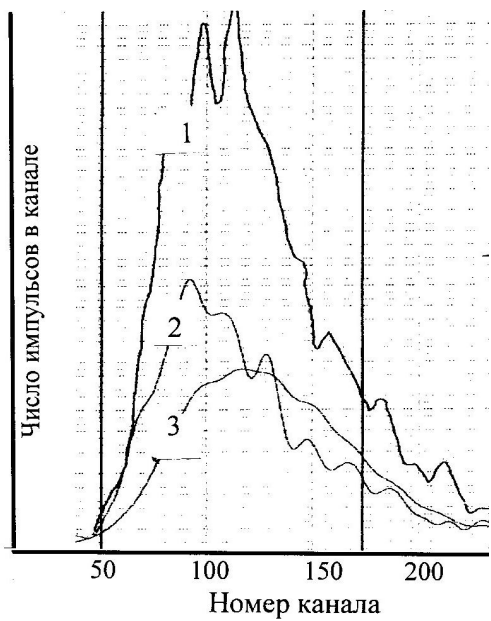

Fig. 2. Apparatus spectra, registered with LSC Quantulus 1220 when measuring the samples of krypton and tritium.

1 − 5 ml sample of krypton; 2 − 5 ml sample of krypton when screening the krypton scintillations with aluminized firm; 3 − spectrum caused by tritium using a liquid scintillator.

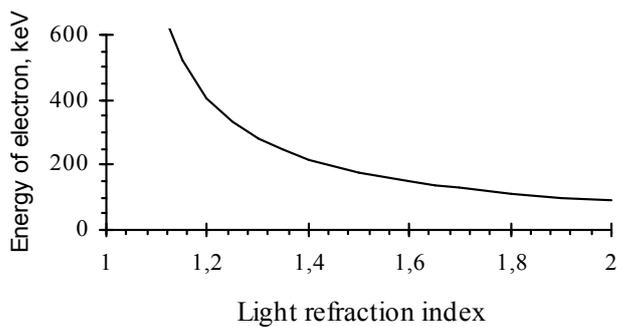

Fig. 3. Dependence of energy threshold of electron required for the generation of Cherenkov radiation in the medium on the light refraction index.